\documentclass[aps,tightenlines,nofootinbib,preprint,groupedaddress]{revtex4}
\usepackage{graphicx}

\newcommand{\be}{\begin{equation}}
\newcommand{\ee}{\end{equation}}
\newcommand{\bear}{\begin{eqnarray}}
\newcommand{\eear}{\end{eqnarray}}
\begin{document}

\title{Heavy quark mass determination from the quarkonium 
ground state energy: a pole mass approach}

\author{Taekoon Lee}
\email{tlee@phya.snu.ac.kr}

\affiliation{Department of Physics, Seoul National University, Seoul 151-742,
Korea}
\affiliation{
National Center for Theoretical Sciences, National Tsing Hua University,
Hsinchu, Taiwan}


\begin{abstract}
The heavy quark pole mass in perturbation theory suffers from a
renormalon caused, inherent uncertainty of $O(\Lambda_{\rm QCD})$.
This fundamental difficulty of determining the pole mass to an accuracy 
better than the inherent uncertainty can be overcome by  direct 
resummation of the first infrared renormalon. We show  how a properly
defined pole mass as well as the $\overline {\rm MS}$ mass for the top
and bottom quarks can be determined accurately from the $O(m\alpha_s^5)$
quarkonium ground state energy. 

\end{abstract}

\pacs{}


\maketitle

\section{\label{sec1}Introduction}
 
The  pole mass is a central parameter in heavy quark physics but suffers from
a renormalon caused ambiguity \cite{bigi,beneke-braun}. In the heavy quark
effective theory (HQET)
it is  defined formally  as the heavy quark
mass  at which the residual mass term vanishes.
This definition of the pole mass is however incomplete.
In general, the parameters and the coefficient functions of HQET should be 
related to the fundamental quantum chromodynamics (QCD)
Lagrangian through the matching process.
In particular, in perturbation theory the pole mass is related to the 
high energy mass such as the $\overline{\rm MS}$ mass
$ m_{\overline{\rm MS}}$ 
($\equiv m_{\overline{\rm MS}}(m_{\overline{\rm MS}}))$ by
\bear
m_{\rm pole}= m_{\overline{\rm MS}}( 
1+\sum_{n=0}^\infty p_n \bar\alpha_s^{n+1})\,,
\label{massexpansion}
\eear
where $\bar\alpha_s=\alpha_s(m_{\overline{\rm MS}})$ denotes the strong
coupling.
The problem with this expansion is that it is an asymptotic expansion,
with a large order behavior \cite{ben}, 
\be
p_n=C_m \frac{(n+\nu)!}{\nu !} (2 \beta_0)^n \left[1+ O(1/n)\right]\,,
\label{largeorder}
\ee
where $C_m$ is a normalization constant and 
$\nu=\beta_1/2\beta_0^2$, with $\beta_0, \beta_1$ denoting the first two
coefficients of the QCD beta function. Because of the factorial divergence
the expansion (\ref{massexpansion}) contains an intrinsic uncertainty
of order
\bear
m_{\overline{\rm MS}}\, \exp{\left(-\frac{1}{2\beta_0\bar\alpha_s}\right)} 
\propto
\Lambda_{\rm QCD}\,,
\eear
which means  that within the  perturbation theory the pole mass cannot
be determined to an accuracy better than $O(\Lambda_{\rm QCD})$ even if
the $\overline{\rm MS}$ mass is known.

On the other hand, one may try to extract the pole mass from the predictions
of physical observables in HQET. But, as well known, this too does not 
work since
the predictions of HQET as well  suffer from the renormalon caused uncertainty.
As an example, consider the inclusive semileptonic $B$ decay rate
in HQET,
\be
 \Gamma(B\to X_{ue\bar\nu_e}) =\frac{G_{\rm F}^2|V_{\rm ub}|^2}{192\pi^3}
 (m^{(b)}_{\rm pole})^5 f(\alpha_s)\{1+ O[1/(m^{(b)}_{\rm pole})^2]\}\,.
\label{decayrate}
\ee
$f(\alpha_s)$ contains the QCD corrections to the
free $b$ quark decay and can be expanded in power series of
the strong coupling. This expansion also suffers from the factorially 
growing divergence similar
to that  in Eq. (\ref{largeorder}) and 
has an intrinsic uncertainty of
 $O(\Lambda_{\rm QCD}/m^{(b)}_{\rm pole})$ \cite{bigi,bbz}.
Because of this uncertainty the bottom quark pole mass $m^{(b)}_{\rm pole}$
cannot be extracted to an accuracy better than $O(\Lambda_{\rm QCD})$,
even assuming  the decay rate is measured precisely and the
Cabibbo-Kobayashi-Maskawa matrix element $V_{\rm ub}$ is known.

Formally, this problem can be resolved by Borel resummation of the
divergent series. Consider again the pole mass expansion 
(\ref{massexpansion}).  The resummed pole mass can
be written as 
\be
m_{\rm pole}= m_{\rm BR} \pm i \Gamma_m\,,
\ee
where $m_{\rm BR}$ and $\Gamma_m$
denote the real and the imaginary parts
 of the Borel integral, respectively. The imaginary part arises from the
renormalon singularity that
gives rise to the large order behavior (\ref{largeorder}), and is ambiguous;
its sign depends on the integration contour taken. 
In general, an unphysical, ambiguous imaginary part in Borel resummation
implies presence of nonperturbative effect and is supposed to be canceled 
by the latter. In the case of the pole mass, however,
the ambiguity is spurious in that in physical observables it is 
canceled by a corresponding ambiguity in another divergent 
series in the coefficient functions or the matrix elements
\cite{neubert,manohr,bbz}. For example, in  the $B$ decay rate
the ambiguous imaginary part in the resummed pole mass
is canceled by the ambiguity in the resummed $f$.
Thus, as far as physical observables are concerned, the introduction of
the nonperturbative effect is not necessary.

Since the ambiguous imaginary parts in Borel resummations in HQET are
canceled in physical observables
we can entirely ignore them and keep only the real parts, the `BR' quantities.
Under this `BR' prescription of Borel resummation the pole mass
is defined as
\be
m_{\rm pole}\equiv m_{\rm BR}\,.
\ee
Throughout the paper we take this as the definition of the pole mass, unless
implied otherwise.
By definition the BR mass, when expanded in the strong coupling constant,
has exactly the same perturbative coefficients as the pole mass in
perturbation theory.

Although this formal argument resolves the renormalon problem
the actual computation of the  `BR' quantities is another matter.
Since the ambiguity in Borel resummation arises from the renormalon 
singularity, to calculate a BR quantity to an accuracy better
than $O(\Lambda_{\rm QCD})$  it is necessary to have a precise description
of the Borel transform  in the region that includes the
origin and the renormalon singularity. Within the perturbation theory 
only the first few terms of the power expansion
of the Borel transform about the origin are known, so describing the
Borel transform beyond the immediate neighborhood of the origin can 
be a difficult task, since it would require a precise 
knowledge on the large order behavior. 

Describing the Borel transform accurately
about the renormalon singularity  is equivalent to having an
accurate information on the associated large order behavior.
Since the essential information in the large order behavior that is missing is
the normalization constant (residue) its computation will make
the computation of BR quantity possible. Fortunately, the normalization
constant can be  calculated in perturbation theory
\cite{lee-residue1,lee-residue2}, and 
the residue of the pole mass expansion can be computed accurately,
within a few percent error, from the known next-next-leading order (NNLO)
calculations of the expansion (\ref{massexpansion})
\cite{lee-pot,lee-decay,pineda}.
This accurate
computation of the pole mass residue is important, because the residues
of other divergent series whose leading IR renormalon is cancelled by that
of the pole mass 
can be obtained to the same accuracy.

Once the residue is known the Borel transform can be expanded systematically
about the renormalon singularity. Then by interpolating 
this expansion with the perturbative expansion about the origin
we can obtain an accurate description of the Borel transform in the
region that contains the origin and the singularity. The interpolation
itself is encoded in what we call bilocal expansion.

This idea of bilocal expansion of the Borel transform was successfully
applied to the heavy quark static potential and the inclusive semileptonic
$B$ decay \cite{lee-pot,lee-decay}.
The Borel resummed heavy quark potential, for example,
at short and intermediate distances agrees remarkably well with lattice 
calculations. 

So far, the renormalon problem in heavy quark physics was handled essentially 
by {\it avoiding} it by only considering physical observables. Since physical
observables are free from renormalons, one can avoid the renormalon
problem by replacing the pole mass with a renormalon-free (short distance)
mass. Instead, the bilocal expansion allows us to confront the renormalon
problem directly, and render us infrared sensitive quantities at our disposal. 

In this paper we apply this technique to
the perturbative expansion of the heavy quarkonium ground state
energy. Our purpose is two fold. First we aim to demonstrate that
the scheme described above indeed allows us to extract a properly 
defined pole mass accurately from experimental data. It is often claimed
that a pole mass cannot be determined to accuracy better than
$O(\Lambda_{\rm QCD})$. We provide here a counterproof.
Secondly, we aim to re-examine the analysis in Ref. \cite{penin}
which extracts from the quarkonium binding energy the top and bottom quark 
masses in the pole mass scheme but
without taking into account the renormalon effect. We show that
a proper handling of the renormalon results in sizable shifts 
in the extracted quark masses. 

\section{\label{method}The method}

In this section we give a brief summary of the  resummation
method of the first infrared (IR) renormalon in heavy quark physics using the 
bilocal expansion. For details we refer the reader to \cite{lee-pot,lee-decay}

Assume a  physical quantity $A(\alpha_s)$ has perturbative expansion
\be
A(\alpha_s)=\sum_{n=0}^\infty a_n \alpha_s^{n+1}\,,
\ee
and a renormalon ambiguity proportional to $\Lambda_{\rm QCD}$.
Then the Borel transform $\tilde A(b)$ defined by the Borel integral,
\be
A(\alpha_s) = \frac{1}{\beta_0} \int_{0\pm i\epsilon}^{\infty\pm i\epsilon}
e^{-b/\beta_0\alpha_s}
\tilde A(b) d b\,,
\label{borelint}
\ee
has the perturbative expansion about the origin,
\be
\tilde A(b) = \sum_{n=0}^\infty \frac{a_n}{n!}
\left(\frac{b}{\beta_0}\right)^n \,,
\label{borelt-o}
\ee
and about the renormalon singularity at $b=1/2$ of the form
\be
\tilde A(b)=\frac{C}{(1-2b)^{1+\nu}} \left( 1+c_1(1-2b) +c_2
(1-2b)^2+\cdots\right) + ({\rm analytic\,\, part})\,,
\label{borelt-s}
\ee
where $C$ and $c_i$ are real constants and $\nu$ same as in
(\ref{largeorder}). The coefficients $c_i$ can be determined
by expanding $\Lambda_{\rm QCD}$ in $\alpha_s$, and they depend only
on the beta function coefficients. In terms of the known four loop beta
function the first two coefficients can be found  as \cite{ben} 
\bear
c_1=\frac{
\beta_1^2-\beta_0\beta_2}{4\nu\beta_0^4}\,,\,\,\,
c_2= \frac{
 \beta_1^4 +4\beta_0^3\beta_1\beta_2
-2 \beta_0\beta_1^2\beta_2
+\beta_0^2(\beta_2^2-2\beta_1^3)-2\beta_3\beta_0^4}{
32\nu(\nu-1)\beta_0^8} \,.
\label{coeffs}
\eear
The ``analytic part'' in (\ref{borelt-s}) denotes
terms analytic on the disk $|b-1/2|<1$ about the singularity.
The radius of convergence of the expansion (\ref{borelt-o}) is expected to be
bounded by the first IR renormalon at $b=1/2$ and that of 
the expansion  (\ref{borelt-s}) bounded by the 
second renormalon at $b=3/2$. The Borel transforms of the pole mass 
and the quarkonium  binding energy are expected to  
satisfy all these conditions.

As noted above, to compute the Borel integral (\ref{borelint}) to an 
accuracy better than
$O(e^{-1/2\beta_0\alpha_s})$ we must have an accurate description
of the Borel transform in the region that includes both the origin and the
first IR renormalon singularity. This can be done with the
bilocal expansion that interpolates the two expansions (\ref{borelt-o}) and
(\ref{borelt-s}),
\bear
\tilde A(b) &=&\lim_{N,M \to \infty} \tilde A_{\rm N,M} (b)\,,
\eear
where
\bear
\tilde A_{\rm N,M} (b) 
&=& 
\sum_{n=0}^N\frac{h_n}{n!} \left(\frac{b}{\beta_0}\right)^n
+\frac{C}{(1-2b)^{1+\nu}}\left[ 1 +\sum_{i=1}^M c_i (1-2b)^i\right]
\label{bilocalexpansion} \,.
\eear
The coefficients $h_n$ are to be determined by demanding
the bilocal expansion have the same perturbative expansion about the
origin as (\ref{borelt-o}).
The first two coefficients $c_{1,2}$ are known, so taking 
$M=2$ in (\ref{bilocalexpansion}) we then have
\bear
h_0&=& a_0 -C(1+c_1+c_2) \,, \nonumber\\
h_1&=&a_1 -2C\beta_0[1-c_2+\nu(1+c_1+c_2)] \,,\nonumber\\
h_2&=& a_2-4C\beta_0^2[2+\nu(3+c_1-c_2)+\nu^2(1+c_1+c_2)]\,,\nonumber\\
h_3&=& a_3-8C\beta_0^3 (1+\nu)[6+\nu(5+2 c_1-c_2)+\nu^2(1+c_1+c_2)]\,,\,\,  
\text{etc.}
\label{h-coeffs}
\eear
The interpolating Borel transform $\tilde A_{\rm N,M}(b)$ implements  the
correct nature of the first renormalon singularity and this
allows us to resum the renormalon caused large order behavior to
all orders. A quick comparison between (\ref{borelt-s}) and
(\ref{bilocalexpansion}) shows that  
the expansion about the origin in (\ref{bilocalexpansion}),
those with the coefficients $h_n$, simulates the 
``analytic part'' in (\ref{borelt-s}).

Clearly, a working bilocal expansion requires computation of 
 the renormalon residue. The residue can be calculated perturbatively 
 in a straightforward manner 
following the observation in \cite{lee-residue1,lee-residue2}.
To compute the residue, notice that
\be
C= R(\frac{1}{2})\,,
\label{residue-c}
\ee
where
\be
R(b)\equiv (1-2b)^{1+\nu} \tilde A(b)\,.
\ee
Now the value of the function $R(b)$ at $b=\frac{1}{2}$ can be evaluated
by a series
expansion in powers of $\frac{1}{2}$, since $R(b)$ is expected to be
analytic on the disk
$|b|< 1/2$ and is bounded at $b=1/2$. Even though the evaluation occurs
precisely on the boundary of the convergence disk, the boundedness of  $R$
on the boundary guarantees convergence.
Note that for the series expansion of $R(b)$
only the Borel transform in  perturbative form (\ref{borelt-o}) is needed.

The convergence of the power expansion for the residue can be improved by
rendering the function $R$ to be smoother on the disk $|b|<1/2$.
This can be done by pushing away from the origin
the ultraviolet renormalons and the subleading IR renormalons through an
employment of a conformal mapping.
For the case of the pole mass and the quarkonium energy we can use the mapping:
\be
w=\frac{\sqrt{1+b}-\sqrt{1-2b/3}}{\sqrt{1+b}+\sqrt{1-2b/3}}\,,
\label{mapping}
\ee
which maps the first IR renormalon to $w=1/5$ and all others onto
the unit circle. In $w$ plane the residue can be obtained by
expanding $R[b(w)]$ about the origin in power series, and evaluating it
at $w=1/5$.

\section{\label{brtoms}
Relation between the pole mass and the $\overline{\rm MS}$ mass}

In this section we show that the relation between the pole mass and
the  $\overline{\rm MS}$ mass in the case of the bottom and top quark
can be given to an accuracy much better than $O(\Lambda_{\rm QCD})$.
Writing the expansion (\ref{massexpansion}) as
\be
m_{\rm pole}= m_{\overline{\rm MS}} [1+{\cal M}(\bar\alpha_s)]\,,
\hspace{.25in}
{\cal M}(\bar\alpha_s)=\sum_{n=0}^{\infty} p_n \bar\alpha_s^{n+1} \,,
\ee
the Borel transform $\tilde {\cal M}(b)$ of ${\cal M}(\bar\alpha_s)$
has the first IR renormalon
singularity precisely of the form (\ref{borelt-s}).

To employ the bilocal expansion for the summation of the
series for ${\cal M}$
we first need the residue $C_m$ for the pole mass, corresponding to $C$
in Eq. (\ref{borelt-s}).
Following the description in the previous section and using the
known first three coefficients \cite{gray,melnikov,chetyrkin}
\bear
p_0&=&0.4244\,,\nonumber\\
p_1&=&1.3621-0.1055 n_f \,,\nonumber\\
p_2&=&6.1404-0.8597 n_f +0.0211 n_f^2\,,
\label{pi}
\eear
where $n_f$ denotes the number of light quark flavors,
we obtain the residue
\be
C_m=\left\{ \begin{array}{l}
    0.4244+0.1151-0.0099 \pm 0.0080=0.5296 \pm 0.0080\,,\,\,(n_f=5)
    \\0.4244+0.1224+0.0101\pm 0.0080=0.5569\pm 0.0080\,,\,\, (n_f=4)
    \end{array}\right.\,.
\label{residue}
\ee
The errors were obtained by taking the differences
between the residues evaluated at NNLO and at next-next-next-leading 
order (NNNLO),
with the latter computed using the NNNLO coefficients estimated 
 following the method in \cite{lee-jeong}.
The good convergence of the residue at such a low order perturbation
is indeed remarkable, and it can be traced to the 
renormalon dominance in the perturbation expansion of the pole mass. 

Using the values for $C_m$ above and the coefficients (\ref{pi})
we can obtain the interpolating Borel transforms $\tilde {\cal M}_{0,2}(b),
\tilde {\cal M}_{1,2}(b),$ and $\tilde {\cal M}_{2,2}(b)$.
With these Borel transforms it is now easy to do the actual
Borel summation. The integration in the Borel integral
can be performed easily in $w$ plane,
defined by (\ref{mapping}), with the contour now along a ray off the
origin to the unit circle in the first (or fourth) quadrant \cite{lee-cvetic}.
Taking the real part of the integral  we have
\be
m_{\rm BR}= \left\{ \begin{array}{ll}
m_{\overline{\rm MS}}( 1+ 0.060336 +0.001310-0.000029 \pm 0.000046)\,,
& (n_f=5\,,\,\, {\rm top \,\,
quark})\\
m_{\overline{\rm MS}}( 1+ 0.1577 +0.0041-0.0003 \pm 0.0001)\,,
& (n_f=4\,,\,\, {\rm bottom \,\, quark}) \end{array}\right.\,.
\label{mass-conversion}
\ee
In this computation we have taken,
for demonstration purpose, $\alpha_s^{[5]}(m_{\overline{\rm MS}})=0.108$ 
and  $\alpha_s^{[4]}(m_{\overline{\rm MS}})=0.22$ for the top and bottom 
quarks, respectively. The errors are from the uncertainties 
in the computed residues in (\ref{residue}).
Notice the remarkable convergence compared to the unresummed series at
the same $\alpha_s$,
\be
m_{\rm pole}= \left\{ \begin{array}{ll}
m_{\overline{\rm MS}}( 1+ 0.0458+0.0097+ 0.0030)\,,
& (n_f=5\,,\,\, {\rm top \,\,
quark})\\
m_{\overline{\rm MS}}( 1+0.093+0.045+0.032)\,,
& (n_f=4\,,\,\, {\rm bottom \,\, quark}) \end{array}\right.\,.
\ee
Taking $m_{\overline{\rm MS}}= 165$ GeV
for the top and $m_{\overline{\rm MS}}=4.2$ GeV for the bottom quark we see 
the uncertainties in the resummed relations are less than 10 MeV
and 1 MeV, respectively ---
much smaller than $\Lambda_{\overline{\rm MS}}$.

\begin{figure}
 \includegraphics[angle=90 , width=10cm
 ]{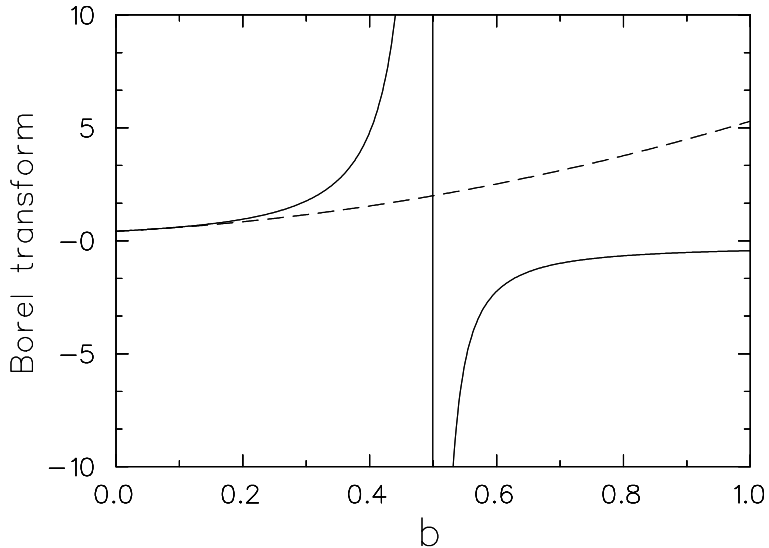}
\caption{\label{fig1} 
Interpolating Borel transform $\tilde {\cal M}_{2,2}(b)$ (solid) vs 
ordinary NNLO perturbative Borel transform (dashed).}
\end{figure}

The drastic improvement in the convergence comes from the proper
treatment of the renormalon singularity in the Borel transform. 
For comparison, we plot in Fig. \ref{fig1}
the ordinary NNLO Borel transform in power
series against the real part of the 
interpolating Borel transform $\tilde {\cal M}_{2,2}(b)$ at $n_f=4$.
Though they agree on the domain about the origin, as should, the deviation
becomes obvious at larger values of $b$, close and beyond the singularity
at $b=1/2$.

The calculation so far clearly shows that direct 
resummation of the first infrared renormalon 
is not only feasible but can be efficient.

\section{Resummation of the binding energy}

Having established a precise connection between the pole mass and 
the $\overline{\rm MS}$ mass we proceed to the resummation of the 
binding energy of the quarkonium $1S$ state. Since we have nothing new to add 
to the nonperturbative effects or the heavy quark finite width effect we 
shall assume the quarks are stable and the system is completely perturbative.
Those effects may be added later on to the perturbative result.

Since the early works by Titard and Yndurain \cite{ty1,ty2} there has been
many studies on the precision calculation of the quarkonium
energy within perturbative QCD. The binding energy is known to NNLO,
and the partial computations of NNNLO energy are also incorporated in
recent works \cite{yuki1,yuki2,pineda,penin}.

The quarkonium $1S$ energy $M_{1S}$ is given as the sum of the pole mass and
the binding energy $E$,
\bear
M_{1S}&=& 2 m_{\rm pole}+ E \nonumber\\
&=&2 m_{\rm BR} +E_{\rm BR}\,.
\eear
In perturbation theory the binding energy can be expanded as
\be
E=- \bar{\mu}(\mu) \sum_{n=0}^\infty {\cal E}_n(\xi) \alpha_s(\mu)^{n+1}\,,
\ee
where $\mu$ is the renormalization  scale and  
\be
\bar{\mu}(\mu)= C_F \alpha_s(\mu) m_{\rm BR}, \hspace{.25in} \xi\equiv
\frac{\mu}{\bar{\mu}}\,,
\ee
and $C_F=4/3$.
The first three coefficients are known \cite{ty1,pineda2}, and, for $m_{\rm
BR}\alpha_s^2 \gg \Lambda_{\rm QCD}$, the fourth  almost
\cite{penin22,basv,kpss,penin}, with only 
the three loop contribution to the static potential not calculated.
They are given by ${\cal E}_n=\frac{1}{3}P_n$ with
\bear
P_0&=& 1, \hspace{.25in} P_1= 4\beta_0L +k_1\,, \nonumber\\
P_2&=& 12\beta_0^2L^2 +(-8\beta_0^2+4\beta_1+6\beta_0k_1)L +k_2\,, \nonumber\\
P_3&=&32 \beta_0^3L^3
+(-56\beta_0^3+28\beta_0\beta_1+24\beta_0^2k_1)L^2 \nonumber\\
&&+(16\beta_0^3-16\beta_0\beta_1+4\beta_2
-12\beta_0^2k_1+6\beta_1k_1+8\beta_0k_2)L
+k_3\,,
\label{coeffs0}
\eear
where $L=\ln(\xi)$ and 
\bear
k_1&=&\frac{1}{\pi}(97/6-11n_f/9)\,,\nonumber\\
k_2&=&\frac{1}{\pi^2}( 337.9471-40.96485 n_f+1.162857 n_f^2)\,,\nonumber\\
k_3&=&\frac{1}{\pi^3}(7078.7900-1215.5475 n_f+
69.450816 n_f^2-1.21475 n_f^3 \nonumber\\
&&+0.031250 \,a_3+474.2893 \ln[\alpha_s(\mu)])\,,
\label{e27}
\eear
and  the beta function coefficients,
\bear
\beta_0&=&\frac{1}{4\pi}(11-2n_f/3),\hspace{.25in}
\beta_1=\frac{1}{(4\pi)^2}(102-38n_f/3)\,,\nonumber\\
\beta_2&=&\frac{1}{(4\pi)^3}(2857/2-5033n_f/18+325n_f^2/54)\,.
\eear
$a_3$ in (\ref{e27}) denotes
the unknown three loop coefficient in the $\overline{\rm MS}$ scheme
of the static potential in momentum space. Using the Borel transform
method of estimating unknown higher order coefficients \cite{lee-jeong},
we can find
an estimate of the three loop coefficient $V_3$ of the static 
potential in coordinates space,
\be
V(r)= \frac{1}{ r} \sum_{n=0}^{\infty} V_n \alpha_s(1/r)^{n+1}\,,
\ee
which reads
\be
V_3=\left\{ \begin{array}{ll} -19.33 \pm 2.73\,, & (n_f=5) \\
                               -27.03 \pm 3.50\,, & (n_f=4)
			       \end{array}\,.\right.
\ee
The corresponding  $a_3$ is then given by \footnote{
This agrees with the existing estimates within the errors,
 $a^3/4^3=60, 98$ from the Pad\`e \cite{alias} and $a^3/4^3=37, 72$ from
 the large order behavior \cite{pineda},
 for $n_f=5,4$, respectively.}
\be
\frac{a_3}{4^3}=\left\{ \begin{array}{ll} 34 \mp 63\,, & (n_f=5) \\
                               59 \mp 81\,, & (n_f=4)
			       \end{array}\,.\right.
\label{vari}
\ee
Note the errors in $V_3$ estimates get amplified in the conversion from
$V_3$ to $a_3$ by an order
of magnitude, resulting in large errors in the latter. So a
few 100\% error in $a_3$ should be counted as normal.

The binding energy has a renormalon caused ambiguity proportional to
$\Lambda_{\rm QCD}$ which is to be canceled by that of the pole mass term
($2 m_{\rm pole}$) \cite{hoang-teub,ben98,rs}, and
so the Borel transform $\tilde E$ of the binding energy can be expanded
about the renormalon singularity as
\be
\tilde E(b,\xi)=-\frac{2C_m \xi\bar\mu}{(1-2b)^{1+\nu}}\left( 1+c_1(1-2b) +c_2
(1-2b)^2+\cdots\right) + ({\rm analytic\,\, part})
\label{eb-sing}
\ee
with $C_m$ and $c_{1,2}$ given by (\ref{residue}) and (\ref{coeffs}),
respectively.
Note that in Eq. (\ref{eb-sing}) the renormalon residue of $\tilde E$
is expressed
in terms of the pole mass residue $C_m$, using the renormalon cancellation
between the pole mass and the binding energy.
This is to utilize the accurate computation of the pole mass residue.
One can, of course, proceed without refereeing to it,
by directly computing the
residue  from the coefficients (\ref{coeffs0}); however, the convergence
is not as good as in the pole mass residue \cite{lee-pot,lee-decay,pineda}.
Combining  (\ref{eb-sing}) with  the expansion about the
origin,
\be
\tilde E(b,\xi)=-\bar\mu \sum_{n=0}^{\infty} \frac{{\cal E}_n(\xi)}{n!}
\left(\frac{b}{\beta_0}\right)^n\,,
\label{eb-org}\ee
into a
bilocal expansion following the description in Sec. \ref{method}
we can obtain, using the known coefficients 
(\ref{coeffs0}), the first four interpolating Borel transforms 
$\tilde E_{N,2}(b,\xi)$ ($N=0,1,2,3$).

The Borel resummed binding energy $E_{\rm BR}$ can then be obtained by
substituting $\tilde E(b,\xi)$ with $\tilde E_{N,2}(b,\xi)$ in
the integral
\be
E_{\rm BR} ={\rm Re}\left[ \frac{1}{\beta_0} 
\int_{0\pm i\epsilon}^{\infty\pm i\epsilon}
e^{-b/\beta_0\alpha_s(\mu)}
\tilde E(b,\xi) d b\right]\,.
\label{e-integral}\ee

In the following we take $\alpha_s^{[5]}(M_Z)=0.1172\pm 0.002$ \cite{pdg},
and use the four loop beta function for the computation of the running
coupling $\alpha_s(\mu)$.
\subsection{Top quark}

Because of the large mass and decay width of the top quark
the toponium $1S$ state is expected to have  small nonperturbative
effects and to be well described by perturbation theory. In future
$e^+e^-$ linear colliders the toponium energy $M_{1S}$ 
is expected to be measured precisely from the peak
position of top threshold production cross section.
This provides a unique
opportunity for precision determination of the top mass.

\begin{figure}
 \includegraphics[angle=90 , width=10cm
 ]{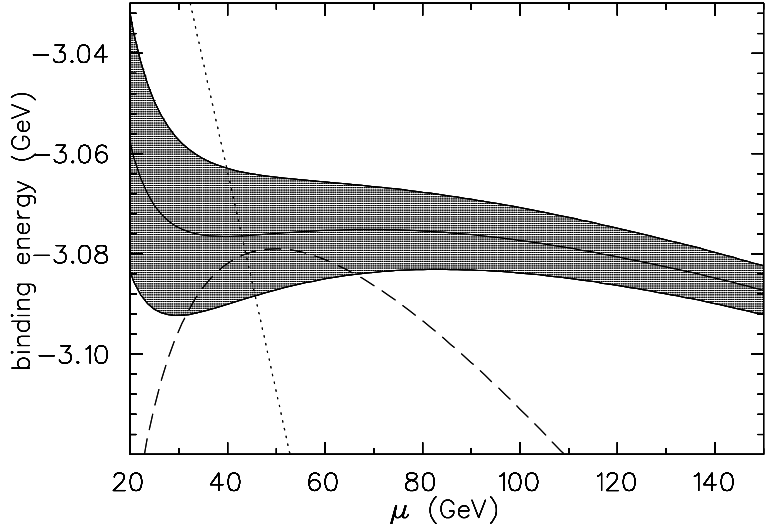}
\caption{\label{fig2} 
The resummed toponium $1S$ state binding energy  using
$\tilde E_{1,2}$(dotted), $\tilde E_{2,2}$ (dashed),
and $\tilde E_{3,2}$ (solid). The shaded band denotes the variation
due to the uncertainty in the estimate of $a_3$ in Eq. (\ref{vari}).
The central line,  upper, and lower boundaries correspond to 
$a_3/4^3= 34, -29,$ and 97, respectively.}
\end{figure}

We now perform the Borel integration in (\ref{e-integral}) with the
interpolating Borel transforms.
The results are plotted in Fig. \ref{fig2} at
the pole mass value $m_{\rm BR}=175\,\, {\rm GeV}\,.$

\begin{figure}
 \includegraphics[angle=90 , width=10cm
 ]{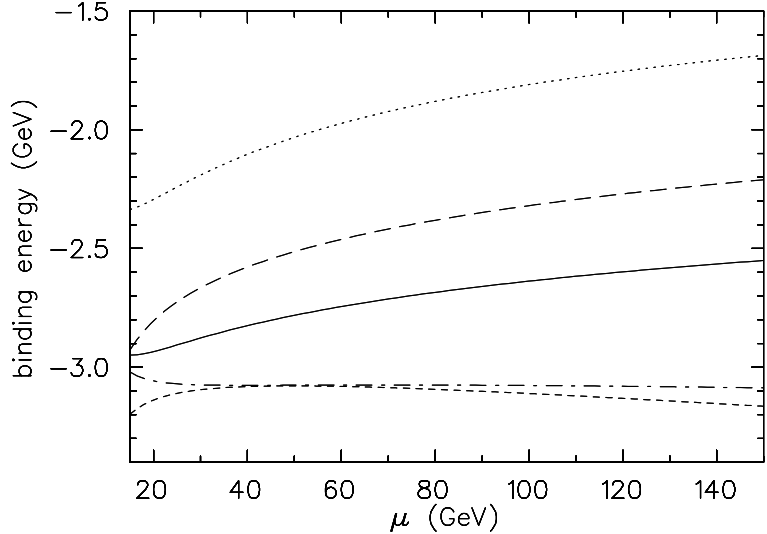}
\caption{\label{fig3} 
The unresummed toponium $1S$ state binding energy at
NLO(dotted), NNLO(dashed),
and NNNLO (solid). For comparison the resummed binding energies using $\tilde
E_{2,2}$  (short-dashed) and $\tilde E_{3,2}$ (dot-dashed) are given. The same
pole mass  was employed in the resummed and unresummed cases.}
\end{figure}

Notice the very small renormalization scale dependence in 
the NNNLO resummed energy.
It is less than $10$ MeV over the range 
$30\,{\rm GeV}\leq \mu\leq 150\,{\rm GeV}$ with $a_3/4^3= 34$.
For comparison we  plot
the unresummed binding energies in Fig. \ref{fig3} using the same pole mass.
Compared to the unresummed binding energies the improvement of the resummed
is clear. They have much smaller perturbation order dependence and
scale dependence. It is evident that for precise 
determination of the binding energy
the renormalon must be taken into account properly. 

The dominant theoretical uncertainty in the resummed binding energy 
comes from the uncertainty in the strong coupling constant.
In the pole mass scheme under consideration there is a natural scale 
for the binding energy,
the Bohr scale $\bar\mu(\mu_{\rm Bohr})=\mu_{\rm Bohr}$, with 
 $\mu_{\rm Bohr}\approx 33$ GeV. We expect an optimal result would come
 around this scale. Thus, taking the variation of the binding energy at the
 Bohr scale under $\delta\alpha_s^{[5]}(M_Z)=0.002$ we
 estimate the error caused by the strong coupling
 to be about $\pm 150$ MeV to the binding energy,
 which would give $\pm 75$ MeV uncertainty
in the pole mass.
Other than the strong coupling the main source of uncertainty is $a_3$, 
as can be seen in Fig. \ref{fig2}.
Comparatively, the renormalization scale dependence is small.
In Fig. \ref{fig2} we see that the binding energies at
next-leading order (NLO), NNLO and NNNLO come close 
together in the range $\mu_{\rm Bohr}\leq\mu\leq
80\,{\rm GeV}$, and  also the principle of minimal sensitivity (PMS)
\cite{pms} scales for the NNLO and NNNLO curves lie
 within the same interval. The renormalization scale dependence over this
 interval of the central NNNLO line  is negligible, less than 2 MeV. The 
 uncertainty in the renormalon residue at $n_f=5$ in (\ref{residue}) 
 causes $\pm 5$ MeV error to the binding energy. The  uncertainty in $a_3$
 causes another  $\pm 16\, {\rm MeV}$ error in the binding energy. 
 Combining these errors we 
 conclude that the current theoretical error, excluding that caused
 by the strong coupling constant, on the pole mass is
  $\pm 10\, {\rm MeV}$.
 
Once the pole mass is extracted from the binding energy
the top mass in the  $\overline{\rm MS}$ scheme can be
determined using the relation between  the 
BR mass and  $\overline{\rm MS}$ mass discussed in Sec. \ref{brtoms}.
The conversion introduces
another error due to the uncertainty in the relation, about $\pm 7$ MeV to 
the $\overline{\rm MS}$ top mass. Combining these uncertainties in quadrature
the corresponding theoretical uncertainty in the $\overline{\rm MS}$  mass
is about $\pm 12\, {\rm MeV}$.
 
\subsection{bottom quark}

The resummation of the $\Upsilon(1S)$ binding energy can proceed 
in the same way as in the toponium, only with a few changes of the parameters.
We take the renormalon residue and $a_3$ estimates in (\ref{residue}) and
(\ref{vari}), respectively,
for $n_f=4$, and for the strong coupling $\alpha_s^{[4]}(\mu)$  we use 
RunDec \cite{rundec} with the four loop beta function and three loop matching.

\begin{figure}
 \includegraphics[angle=90 , width=10cm
 ]{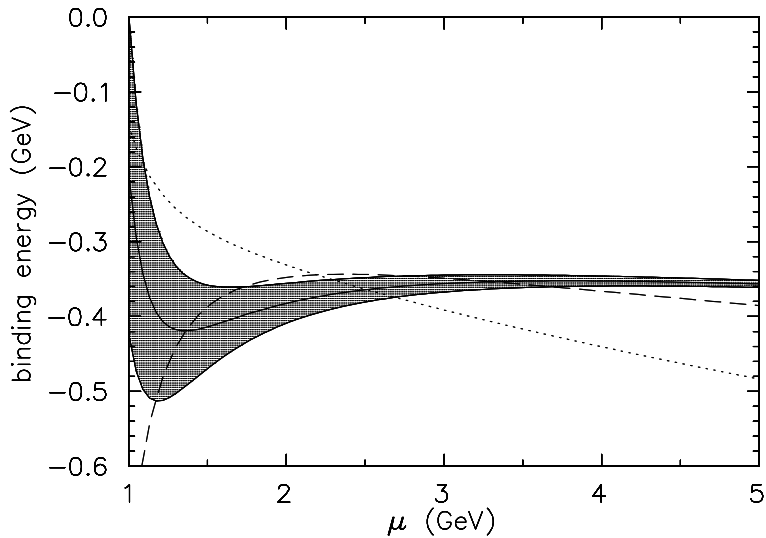}
\caption{\label{fig4} 
The resummed $\Upsilon(1S)$  binding energy  using
$\tilde E_{1,2}$(dotted), $\tilde E_{2,2}$ (dashed),
and $\tilde E_{3,2}$ (solid). The shaded band denotes the variation
due to the uncertainty in the estimate of $a_3$ in Eq. (\ref{vari}).
The central line,  upper and lower boundaries correspond to 
$a_3/4^3= 59, -22,$ and 140, respectively.}
\end{figure}

\begin{figure}
 \includegraphics[angle=90 , width=10cm
 ]{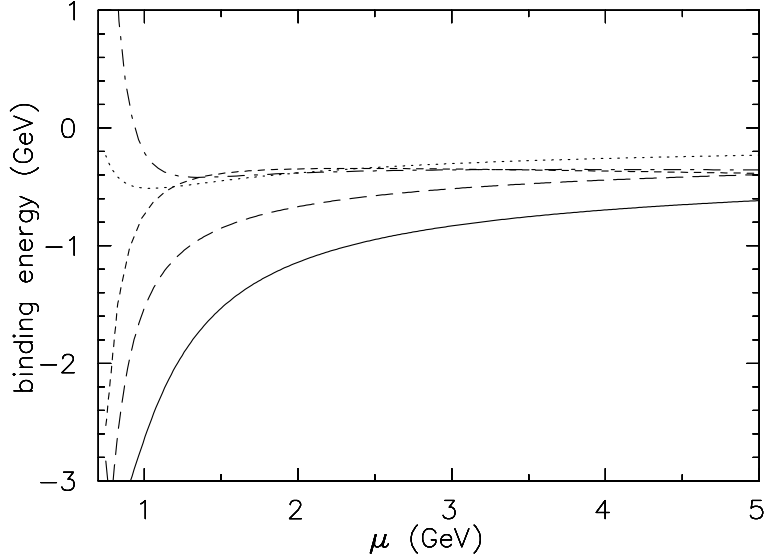}
\caption{\label{fig5} 
The unresummed  $\Upsilon(1S)$  binding energy at
NLO(dotted), NNLO(dashed),
and NNNLO (solid). For comparison the resummed binding energies using $\tilde
E_{2,2}$  (short-dashed) and $\tilde E_{3,2}$ (dot-dashed) are given. The same
pole mass  was employed in the resummed and unresummed cases.}
\end{figure}

The resummed binding energies from the interpolating Borel transforms at
NLO, NNLO, and NNNLO are plotted in Fig. \ref{fig4} at the pole mass
$m_{\rm BR}=4.9$ GeV. For comparison the unresummed binding energies 
are plotted in Fig. \ref{fig5}.
Again, the improvement of the resummed energy is clear. It has better 
convergence and less scale dependence. 

Before the BR mass
can be extracted from the fitting of the resummed energy to the precisely
measured $\Upsilon(1S)$ mass, the nonperturbative effects and the
charm mass effect should be subtracted from the  $\Upsilon(1S)$ mass.
Without taking these effects into account we obtain 
$m_{\rm BR}\approx 4.9$ GeV.
The leading nonperturbative effect comes from the Stark effect of the long
range gluon fields in the QCD vacuum \cite{voloshin,leutweyler}.
When the correlation time of the quarkonium state 
is much smaller than $1/\Lambda_{\rm QCD}$ the nonperturbative effect
may be expressed in terms of local gluon condensate. Whether 
$\Upsilon(1S)$ is such a state is not clear, but assuming so,
the nonperturbative effect is given by
\be
\frac{1872}{1275}\frac{m_{\rm BR} \pi<\alpha_s G_{\mu\nu}^2>}{
(C_F\alpha_s m_{\rm BR})^4}\,.
\label{condensate}
\ee
The gluon condensate is poorly known. In general, fitted power
corrections are contaminated by perturbative corrections, and
sensitive to the perturbation order employed in fitting.
For gluon condensate, it has been
observed \cite{zyablyuk} that with higher order perturbation 
the value extracted is 
much smaller than the original sum rule value in \cite{svz1}.
Since in the binding energy the perturbative correction is already at
relatively high order (NNNLO) \footnote{The perturbative 
ultrasoft correction itself starts at $O(m\alpha_s^5)$ \cite{kniehl}.}
the condensate for Eq. (\ref{condensate}) should not contain 
significant perturbative corrections, and thus likely to be
smaller than the original sum rule value.
Here we take  $<\alpha_s G_{\mu\nu}^2>=0.02\pm 0.02$.
Since the local condensate approximation for the nonperturbative
effect is likely to provide an upper bound for any potential
nonlocal effect \cite{gromes}, it may be reasonable to assume that this value
 covers adequately the nonperturbative effect.
With  $m_{\rm BR}=4.9$ GeV we then get the nonperturbative effect 
$60\pm 60$ MeV in the binding energy.
For the nonzero charm mass effect we
find,  using the result in \cite{hoang,soto},
$-20 \pm 15$ MeV in the binding energy.
Subtracting these two effects from the experimentally measured
$\Upsilon(1S)$ mass and
comparing the subtracted mass with the PMS value of the resummed energy, 
$2m_{\rm BR}+E_{\rm BR}$,
we obtain $m_{\rm BR}=4.89 \pm 0.03$ GeV. 

For the error estimate we notice in Fig. \ref{fig4} 
the resummed energies at NLO, NNLO, and NNNLO converge 
around $\mu\approx 2.5$ GeV and the 
PMS scales for the NNLO and NNNLO resummed energies lie within the
range $m_{\rm Bohr}\leq\mu\leq 5$ GeV, where $m_{\rm Bohr}\approx 2$ GeV.
We therefore take the variations of the resummed energies in this
range as the error estimate. Then the uncertainties in the $a_3$ and the
strong coupling give  $\pm 27$ MeV and $\pm 13$ MeV, respectively, to
the binding energy, and the renormalization scale 
dependence and the uncertainty in the residue
(\ref{residue}) cause another $\pm 30$ MeV and  $\pm 12$ MeV, respectively.
The error due to the unknown higher order contribution should be partly covered
by the error from the residue. However, setting this fact aside we shall
assign an independent error $\pm 35$ MeV to the binding energy, which is
the difference between the respective energies at NNLO and at NNNLO at
the Bohr scale. It is interesting to observe that the PMS values at
NNLO and NNNLO are virtually identical, an indication that the unknown higher
order effect should be small.

Combining these errors (including those from the gluon condensate and the
nonzero charm mass effect) in quadrature we have the BR mass for the 
bottom quark
\be
m_{\rm BR}^{(b)}=4.89 \pm 0.04 \,\,{\rm GeV}\,.
\ee
Now, from the BR mass the $\overline{\rm MS}$ mass can be obtained through the
relation, $m_{\overline {\rm MS}}^{(b)}\{1+ {\cal M}_{\rm
BR}[\alpha_s^{[4]}(m_{\overline {\rm MS}}^{(b)})]\}=
m_{\rm BR}^{(b)}$, from which 
we obtain
\be
m_{\overline {\rm MS}}^{(b)}= 4.20 \pm 0.04 \,\, {\rm GeV}\,.
\label{bmass1}
\ee
Compare this to the result from Ref. \cite{penin},
\be
m_{\overline {\rm MS}}^{(b)}= 4.346 \pm 0.070 \,\, {\rm GeV}\,,
\label{bmass2}
\ee
which was obtained without taking the renormalon into account.
Notice that the central values  have a significant difference, and
our value has a smaller error.
The  reduction in the error largely  comes from the
smaller renormalization scale dependence. The assigned error for the
scale dependence in (\ref{bmass2}) is more than four times 
larger than ours, because
of much stronger scale dependence in its derivation.
This comparison shows that renormalon effect is significant and must be
taken into account for accurate determination of the bottom quark mass.
Our value for the bottom quark mass (\ref{bmass1}) compares favorably with
the values collected in \cite{luke-aida}.

\section{Discussion}
In this paper we have shown that 
an accurate calculation of the normalization
constant of the large order behavior and information on the
renormalon singularity allows us a precision calculation of 
the pole mass and quarkonium binding energy. The existing method of handling
the renormalon problem was bypassing it, by concentrating
exclusively on directly measurable observables and employing renormalon-free
short distance masses. In contrast, in our approach each of the first 
IR renormalon in IR sensitive quantities is resummed independently,
and as a consequence those IR sensitive quantities can be extracted
accurately from experimental data.

One of the advantages of our approach is that it provides a natural
solution to the scale mixing problem 
of the  approaches based on renormalon
cancellation. In systems with far-separated multiple scales the
renormalon cancellation approach inevitably mixes those scales through the
renormalization scale, forcing one to choose an optimal scale over a
wide range of scales. Depending on the problem this can be a source
of significant error. For instance, in the top-pair threshold production
the normalization of the production cross section is known to be
sensitive to the
choice of the renormalization scale \cite{toppair},
which can limit the accuracy of the
extracted strong coupling constant. With our approach,
this problem can be solved by resumming the top pole mass and 
the interquark potential at their respective optimal scales,
namely, the top mass and the interquark distance.

This suggests a potential improvement in the bottom quark
mass determination. Instead of applying the resummation
to the perturbatively calculated binding energy, as we did in this paper,
one may solve the Schroedinger equation, which was
advocated in \cite{shroe}, with a resummed interquark 
potential to obtain a more accurate binding energy. It should be noted,
however, presently the error in the bottom quark mass is dominated by
the nonperturbative effect.

\begin{acknowledgements}
The author wish to thank A. Pineda for patiently answering questions.
He also thanks Kingman Cheung and high energy theory group at NCTS, Hsinchu,
Taiwan, for warm hospitality extended to the author during his visit.
This work was supported partly by the BK21 program.

\end{acknowledgements}


\bibliographystyle{JHEP}
\bibliography{polemass}

\end{document}